\documentclass[aps,prl,twocolumn,showpacs]{revtex4}

\usepackage{amsmath}
\usepackage{graphicx}
\usepackage{epstopdf}

\newcommand{\GG}{\mathcal{G}}
\newcommand{\tJ}{\tilde{J}}
\newcommand{\gmix}{g_{\mathrm{mix}}}
\newcommand{\tgmix}{\tilde{g}_{\mathrm{mix}}}

\begin{document}

\title{A thermodynamic approach to phase coexistence in ternary
  phospholipid-cholesterol  mixtures}   
\author{J. Wolff, C. M. Marques, F. Thalmann$^{\ast}$}
\affiliation{Institut Charles Sadron, Universit\'e de Strasbourg, CNRS UPR
22, 23 rue du Loess, Strasbourg Cedex, F-67037, France.}
\date{\today}

\begin{abstract}
We introduce a simple and predictive model for determining the phase stability of ternary phospholipid-cholesterol mixtures. Assuming that competition between the liquid and gel order of the phospholipids is the main driving force behind lipid segregation, we derive a Gibbs free-energy of mixing, based on the thermodynamic properties of the
lipids main transition. A numerical approach wax devised that enable the fast and efficient determination of the ternary diagrams associated with our Gibbs free-energy. The computed phase coexistence diagram of DOPC/DPPC/cholesterol reproduces well known features for this system at 10$^{\circ}$C, as well as its evolution with temperature.
\end{abstract}

\maketitle

%
%
Lipid membranes have captured the attention of physicists, chemists and biologists alike owing to their prominent structural and functional role in living cells, and to the pervading use of lipid based products in the pharmaceutical, food and cosmetic industries~\cite{Mouritsen_MatterOfFat,Heimburg_BiophysicsMembrane}. In the recent years, ternary lipid systems comprising phospholipids and cholesterol were established as important model systems in relation with lateral lipid segregation phenomena in membranes. At the root of this rich phenomenology lies a complex phase behavior of certain lipid mixtures, in which cholesterol molecules play a unique role. Ternary systems comprising a phospholipid in the liquid state, a second phospholipid in the gel state and cholesterol display several unique characteristics, the two most important of them being arguably a wide binary coexistence gap between the so-called liquid disordered ($L_d$) and liquid ordered ($L_o$) phases, and a region of triple coexistence~\cite{2007_Zhao_Feigenson,2007_Veatch_Gawrisch,2008_Goni_Thewalt}.

A prominent feature of phospholipid membranes is their main thermodynamic transition, from a low temperature gel or rippled gel phase (usually termed $L_{\beta}$,$P_{\beta}$), to a high temperature liquid phase ($L_{\alpha}$)~\cite{Heimburg_BiophysicsMembrane}.  A sharp change in the statistical conformations of the hydrophobic chains occurs at the transition, associated with membrane thinning and an increase of the area per phospholipid. The transition takes place at a melting temperature $T_m$ whose precise value depends on the chemical structure of the chains. It is a weakly first order transition for membrane of pure composition. The thermotropic behavior of phospholipid binary mixtures have been studied since the seminal work of Mabrey and Sturtevant~\cite{1976_Mabrey_Sturtevant} and supports the view that liquid-prone and gel-prone lipids tend to phase separate. This suggests that mixtures composed of phospholipids with same head group have their miscibility behavior governed by the competition between their liquid and gel conformations. Hence, a description of ternary lipid systems must necessarily, and to some extent, account for the internal state of the phospholipids.

We introduce in this Letter a simple, robust and predictive model of phospholipid-cholesterol mixtures phase coexistence. Our model aims at predicting the miscibility behavior of phospholipids, based upon the thermodynamic properties of their main transition~\cite{1980_Priest,1983_Sugar_Monticelli,2005_Komura_Olmsted,2008_Putzel_Schick}. The thermotropic behavior of the phospholipid tails is modeled with an internal order parameter, in a way that schematically reproduces the thermodynamic features of the liquid-gel transition.  The peculiar
features of cholesterol-phospholipid interactions~\cite{1999_Radhakrishnan_McConnell,2005_Radhakrishnan_McConnell} are accounted for by means of a non-ideal mixing entropy generalizing the one proposed by Komura \textit{et al.}~\cite{2004_Komura_Andelman}.  In addition, we outline an original numerical procedure for the derivation of ternary phase diagrams.

The stability of a mixture is governed by its Gibbs free-energy of mixing $G_{\mathrm{mix}}$. This thermodynamic function expresses the free-energy difference between the homogeneous mixed state and the system comprising the separated pure components with same global composition. Ternary mixtures in membranes are therefore described by $G_{\mathrm{mix}}(T,n_1,n_2,n_3)$, where the $n_i$ are the molar content of each lipid, and $T$ the temperature. We assume that the hydrostatic pressure remains constant, close to the standard value, and that lipid membranes have no tension, so that these two intensive parameters are absent from $G_{\mathrm{mix}}$. Membranes are assumed to be well hydrated with a large excess of water, and free from any other  external fields~\cite{2008_rozycki_lipowsky}.

The simplest realization of our model, described in this Letter, assumes that the areas per lipid $a_i$ do not change with temperature. This allows us to express $G_{\mathrm{mix}}$ in terms of the surface fractions $\phi_1$, $\phi_2$ and $\phi_C$ of each lipid component, $\phi_C$ standing for the cholesterol. We claim that it is acceptable to neglect all changes in the area per lipid at $T=T_m$, in a first approximation to the description of the mixture phase coexistence, especially in the absence of membrane tension. This restriction is not a fundamental one, and versions of our model accounting for membrane expansion at the transition will be presented elsewhere~\cite{2010_Wolff_toCome}.

A frequent approximation to non-ideal solutions is obtained by combining an ideal mixing entropy with a quadratic enthalpic term. For a membrane of area $\mathcal{A}$, the free-energy of mixing per unit of area, $\gmix=\beta G_{\mathrm{mix}}/\mathcal{A}$, reads: 
\begin{eqnarray}
\gmix &=&
\frac{\phi_1}{a_1}\ln(\phi_1)+\frac{\phi_2}{a_2}\ln(\phi_2)
+\frac{\phi_C}{a_C}\ln(\phi_C) \nonumber\\ 
& & +\chi_{12}\,\phi_1\phi_2+\chi_{1C}\,\phi_1\phi_C
+\chi_{2C}\,\phi_2\phi_C, \label{eq:FH}  
\end{eqnarray}
with $\phi_1+\phi_2+\phi_C=1$ and $\beta$ the Boltzmann factor $1/k_BT$, and $a_1$, $a_2$ and $a_C$ are the areas of the three different lipid species. One observes that expression~(\ref{eq:FH}) alone does not account for the main transition, and that the effective interaction parameters $\chi_{ij}$ are expected to display a non trivial and unknown temperature dependence. We could not obtain, based on~(\ref{eq:FH}), any diagram comparable with the experimental ones. Moreover, our attempts to do so lead to unexpectedly large values of $a_C$, conflicting with the view that cholesterol molecules are smaller than phospholipids.

Experiments and simulations either suggest that cholesterol may form transient bound complexes with phospholipid moieties surrounding them~\cite{1999_Radhakrishnan_McConnell,2005_Radhakrishnan_McConnell}. The presence of such complexes prevents cholesterol molecules from occupying too many neighboring positions, thus giving the semblance of a large excluded volume. To account for the statistical behavior of cholesterol molecules, we derived the following entropic contribution, where all lipid areas have been set to unit:
\begin{multline}
{\GG}^{(E)}(\phi_1,\phi_2)=\phi_1\ln(\phi_1)
+\phi_2\ln(\phi_2)+\phi_C\ln(\phi_C)\\
-(\phi_1+\phi_2)\log(\phi_1+\phi_2)
+(\delta^{-1}-\phi_C)\log(1-\delta\phi_C).
\label{eq:CholEnt}
\end{multline}
This expression originates from a lattice enumeration of configurations, following Flory, where each cholesterol molecule deprives an average number $\delta$ of neighboring sites from the possibility of being occupied by another cholesterol molecule. Expression~(\ref{eq:CholEnt}) reduces to ideal mixing when $\delta\to 1$, and is similar to the entropy of Komura \textit{et al.}  for binary systems and $\delta=2$~\cite{2004_Komura_Andelman}. It accounts for both the mixing term and the excluded volume between cholesterol molecules, consistent with the presence of cholesterol complexes. It finally sets a maximal value $\phi_{C}\leq 1/\delta$ to the mixture content in cholesterol.

Several Landau order parameters have been proposed for describing the liquid-gel chain melting~\cite{1980_Priest,1983_Sugar_Monticelli, 1989_Goldstein_Leibler,2004_Komura_Andelman,2008_Putzel_Schick}. Theories based on a single scalar order parameter can pretend to capture only rough features of the main transition. This is why, for simplicity and generality, we restrict ourselves to a schematic approach, a two-state model based on a scalar order parameter $m$ restricted to the interval $[-1,1]$. The mean-field Ising model under finite external field serves as a guide~\cite{Chaikin_Lubensky}.
\begin{multline}
{\GG}^{(I)}(T,\phi_1,\phi_2,m)= -m \{h_1(T)\phi_1+
h_2(T)\phi_2\} -2\tJ m^2 \\ 
+\large(\frac{1+m}{2}\large)\ln\large(\frac{1+m}{2}\large) 
+\large(\frac{1-m}{2}\large)\ln\large(\frac{1-m}{2}\large),
\label{eq:IsingG}
\end{multline}
with
\begin{equation}
h_1(T)=\frac{\Delta H_1}{2RT_1^2}(T-T_1)\,;\,
h_2(T)=\frac{\Delta H_2}{2RT_2^2}(T-T_2),
\label{eq:effectiveField}
\end{equation}
and $R$ is the gas constant. Eq.~(\ref{eq:IsingG}) represents a mean-field Ising model in the presence of an effective external field $h_{\mathrm{eff}}=h_1\phi_1+h_2\phi_2$. For $\tJ\geq \tJ_c=1/4$, the $m$ dependence of $\GG^{(I)}$ is shaped as a double well potential, with a linear bias induced by $h_{\mathrm{eff}}$. The order parameter $m$ fluctuates freely around an optimal value $m^{\star}$ located at the global minimum of $\GG^{(I)}$, with $(T,\phi_1,\phi_2)$ constant. In practice, as soon as $\tJ$ is larger than 0.3, the two local minima are close to $\pm 1$. We conventionally ascribe $m_g\simeq -1$ to represent the $L_{\beta}$ gel phase, while $m_l\simeq 1$ is set to represent the $L_{\alpha}$ liquid phase. The coupling $\tJ$ expresses that neighboring lipid tails tend to be in the same state, irrespective of the chemical species to which they belong. This is the main cooperative effect that induces the liquid-gel separation.

\begin{figure}
\resizebox{0.45\textwidth}{!}{\includegraphics*{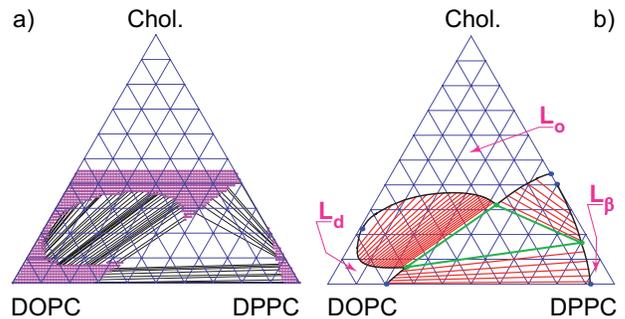}}
\caption{(Color online) 
(a) Discrete approximation to the ternary phase diagram obtained from the lower convex hull method explained in the text.
  (b) Final phase diagram determined from the analytical refinement of the discrete approximation.
}%
\label{fig:method}
\end{figure}


The term $h_{\mathrm{eff}}$ acts as a switch that drives the order parameter from its $m^{\star}=m_g$ gel state to its $m^{\star}=m_l$ liquid state, upon changing its sign from negative to positive. One sees from~(\ref{eq:effectiveField}) that if lipid~1 is pure, $h_1(T)$ changes its sign at $T=T_1$ while $m^{\star}(T)$ suddenly jumps from $m_g$ to $m_l$, with an amplitude $\Delta m^{\star}\simeq 2$. The enthalpic change is close to $\Delta H_1$, so that $\Delta H_1$ can be identified with the molar latent heat at the main transition. The difference in Gibbs free-energy between the liquid and gel phases at $T\neq T_1$ is then extrapolated linearly with $T$ at the vicinity of the transition.  The contribution $\GG^{(I)}$ is thus entirely specified by the temperature and the enthalpy difference at the transition of the lipid species composing the mixture. This term induces a liquid-gel separation in the whole temperature range $[T_1,T_2]$.

Cholesterol-phospholipid interactions are further accounted for with a specific interaction term 
\begin{equation}
{\GG}^{(C)}(\phi_1,\phi_2,m) = \chi\phi_1\phi_C
  -\xi m(\phi_1+\phi_2)\phi_C
\label{eq:CholInt}
\end{equation}
Eq.~(\ref{eq:CholInt}) comprises a $m$-dependent contribution, proportional to $\xi$, which marks the preference of cholesterol for either the $m=1$ liquid order, or the $m=-1$ gel order, according to the sign of $\xi$. This interaction does not discriminate between saturated and unsaturated lipids, since it describes a coupling to the chain order. Other cholesterol-phospholipid interactions are accounted for by a $m$-independent contribution in $\chi$. This order-insensitive interaction introduces a coupling between cholesterol and the unsaturated component. We found that interactions between cholesterol and the saturated component are not required to describe the experimental phase diagrams.

The sum $\GG(T,\phi_1,\phi_2,m) = {\GG}^{(E)}+{\GG}^{(I)}+{\GG}^{(C)}$ is our main thermodynamic potential. In the absence of surface tension, and assuming that changes in internal state occur at constant volume, $m$ fluctuates around its optimal value $m^{\star}$, with $m=m^{\star}$ in the thermodynamic limit.  Therefore, the free-energy of mixing $\gmix(T,\phi_1,\phi_2)$ is obtained by minimizing $\GG(T,\phi_1,\phi_2,m)$ with respect to $m$ at fixed temperature and surface fractions~\cite{2008_Putzel_Schick}.

\begin{eqnarray}
\gmix(T,\phi_1,\phi_2)&=&\min_{m}
\left.\left[\GG(T,\phi_1,\phi_2,m)\right]\right|_{T,\phi_1,\phi_2}\nonumber\\
 &=& \GG(T,\phi_1,\phi_2,m^{\star}(T,\phi_1,\phi_2)).
\label{eq:totalFE}
\end{eqnarray}

The Gibbs free-energy surface $\gmix$ arises from projecting out $m$ from the four-dimensional potential $\GG$. Equivalently, $\gmix$ can be viewed as the restriction of $\GG$ to the submanifold $\partial_m\GG=0$. For $\tJ>\tJ_c$, the constraint $\partial_m\GG=0$ defines two stable submanifolds, a negative $m_g(T,\phi_1,\phi_2)$ and a positive $m_l(T,\phi_1,\phi_2)$ branch, linked respectively with the gel and liquid phases. At fixed $(T,\phi_1,\phi_2)$, only one of the two branches is the stable one, while the other branch remains metastable.

In ternary mixtures it is known that phase coexistence is dictated by the convexity properties of $\gmix(T,\phi_1,\phi_2)$ relative to $\phi_1,\phi_2$. Minimization of $\gmix$ is equivalent to finding $\tgmix$, the lower convex hull of $\gmix$, that displays regions of three kinds. Points of the convex hull in contact with the original surface are stable monophasic regions. Regions lying above triple tangent planes corresponds to triple coexistence. The remaining part of the convex hull correspond to developable patches of surface, wrapped around double tangent lines that connects pairs of points with distinct composition (tie-lines)~\cite{Gibbs}. In practice, we used a
public domain routine, \textit{qhull}, to compute a discrete approximation of $\tgmix$ ~\cite{1996_Barber_Huhdanpaa}. The starting point is a fine mesh discretization of $\gmix$ over the relevant domain of composition $\lbrace\phi_1,\phi_2\geq 0,\phi_1+\phi_2\leq 1-1/\delta \rbrace$. Then \textit{qhull} computes a triangulated surface approaching $\tgmix$. Triple coexistence correspond to facets with all sides much larger than the initial mesh size. Double coexistence is associated with elongated triangles, with a shortest side much smaller than the two longer sides, the latter being oriented parallel with the tie-lines. Finally, small facets of the convex minimization are linked with stable monophasic regions. The projection of the triangulated $\tgmix$ surface onto the composition plane provides directly a fast and accurate approximation of the Gibbs phase diagram shown in fig.~\ref{fig:method}(a). This discrete solution can then be refined according to the usual rules for the determination of ternary phase diagrams.  Details of these calculations will be given in a forthcoming publication~\cite{2010_Wolff_toCome}.

\begin{figure}
\resizebox{0.45\textwidth}{!}{\includegraphics*{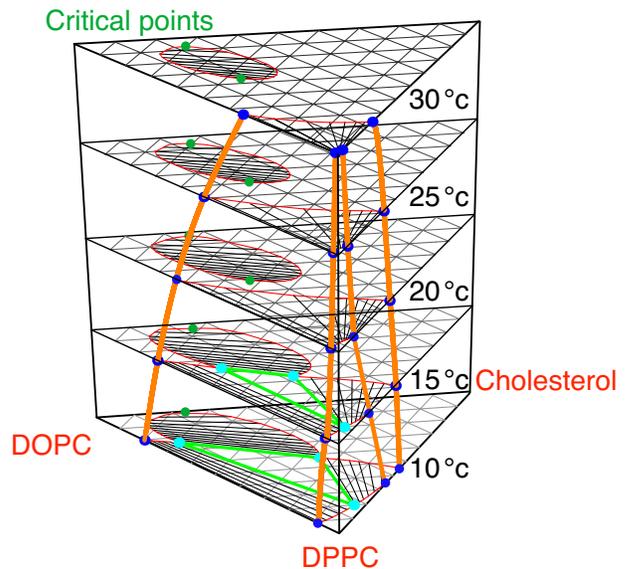}}
\caption{(Color online)
Computed evolution of the DPPC/DOPC/Chol triangular diagram with
  temperature. The computed diagrams display the main expected features for this lipid mixtures. }
\label{fig:prism}
\end{figure}

Arguably, the most widely studied ternary lipid phase diagram for membranes containing cholesterol is the ternary mixture DOPC/DPPC/Chol. DOPC is a double tail molecule with $18$ carbons per tail and a \textit{cis}-unsaturated bond on the ninth carbon. It has a low main transition temperature $T_1$ at $-21^{\circ}$C and an associated enthalpy change $\Delta H_1 $ evaluated at $~7.7$ kcal/mol~\cite{Cevc}. DPPC has also a phosphatidylcholine head but here the two $16$-carbon tails have no unsaturated bonds, leading to a higher transition temperature $T_2$ at $42^{\circ}$C, and associated enthalpy change $\Delta H_2=8.7$ kcal/mol~\cite{Cevc}. Within our approach, the above temperature and enthalpy values and a single coupling parameter $\tJ=0.35$ - see eq.~(\ref{eq:IsingG}) - describe well published data~\cite{2009_Schmidt_Davis}  for the experimental binary DOPC/DPPC mixing behavior. For a complete description of the ternary phase diagram our model further requires three parameters: $\delta,\xi$ and $\chi$. The value of $\delta$ is bound by the maximum amount of cholesterol that a membrane  can hold, estimated here at $\phi_{C}\simeq 0.45 $ or equivalently $\delta=1/\phi_{C}=2.2$~\cite{2007_Zhao_Feigenson}. The parameter $\xi$, as defined in eq.(\ref{eq:CholInt}),  quantifies the preference of cholesterol for the liquid disordered phase with respect to the gel phase~\cite{Mouritsen_MatterOfFat,1987_Ipsen_Zuckermann}.
Binary diagrams of DPPC/Chol computed with  $\xi=1.8$ reproduce well the temperature behavior of the miscibility gaps occurring between the $L_\beta$-gel phase and the $L_o$ liquid ordered phase as determined from Giant Unilamellar Vesicles in \cite{2003_Veatch_Keller} and from NMR~\cite{1990_vist_davis} - see fig.(\ref{fig:prism}). Eq.(\ref{eq:CholInt}) also includes a repulsive $m$-independent  interaction between  cholesterol and the unsaturated component, given by $\chi$. At $10^\circ$C a value of $\chi=6.0$ promotes both a triple coexistence region and a large biphasic domain, with oblique $L_o$/$L_d$ tie lines of positive slope and  a plait point on the left side of the Gibbs triangle, in agreement with experimental findings~\cite{2008_Goni_Thewalt}.  A larger value of $\chi$ would open there a miscibility gap, unseen for DOPC/Chol, while a  smaller value would drastically reduce the biphasic $L_o$/$L_d$ domain. In our model a continuous path connects the $L_o$ and $L_d$ regions  around the critical point, where the order parameter $m$ remains positive. The $L_\beta$ phase on the lower right corner of the phase diagram is associated with a negative $m$-value and separated from the rest of the diagram by a discontinuity in $m$ across the coexistence gap. 
 
The evolution of the Gibbs diagram with respect to temperature is displayed in fig.~(\ref{fig:prism}).  The diagram was drawn here under the assumption of a constant $\xi$ parameter, and a constant value of the coupling contribution $k_B T\cdot \chi $ denoting its pure enthalpic nature. Our diagram at the  lower temperature in fig.~(\ref{fig:prism}), displays the main expected features for a DOPC/DPPC/Chol mixture. As the temperature increases, the triple coexistence triangle shrinks and vanishes between $15^{\circ}$C and $20^{\circ}$C. Also, the $L_o$/$L_d$ biphasic region, resting initially on the left side of the triangle, detaches and assumes a closed shape of decreasing surface. A first critical point, on the left side of the Gibbs diagram, evolves slowly away from the edge~\cite{2007_Veatch_Gawrisch}. A second critical point, initially hidden by the triple coexistence region, is predicted to emerge at the binary detachment temperature. Note also that within our approach the extension of the $L_\beta$ domain, on the right bottom part of the diagram, gradually decreases up to $T_2$.

As a summary we presented a model for phase coexistence in ternary phospholipid-cholesterol mixtures. Our approach is based on a mean-field two-state description of the phospholipids and a non-ideal mixing entropy for cholesterol. All but one parameter in our system are bound by known thermodynamic or physical properties. We combined analytical theory and numerical convex minimization to show the predictive power of our model for the ternary mixture DOPC/DPPC/Chol. Resulting phase coexistence diagrams reproduce well known features for this system at $10^\circ$C as well as the temperature variation of the diagram. We are also encouraged by preliminary results from related ternary systems that further indicate that our model successfully describes a large variety of lipid/lipid/cholesterol mixtures.

$^{\ast }$fabrice.thalmann@ics-cnrs.unistra.fr



\end{document}